\documentclass[aps,prl,superscriptaddress,showpacs,preprintnumbers,amsmath,amssymb]{revtex4}

\usepackage{graphicx}
\usepackage{epsfig}
\usepackage{dcolumn}
\usepackage{bm}

\def\fb{\mbox{fb$^{-1}$}}
\def\acp{\mbox{${\cal A}_{CP}$}}
\def\bb{\mbox{$B\overline{B}$}}
\def\qq{\mbox{$q\overline{q}$}}
\def\mbc{\mbox{$m_{bc}$}}
\def\de{\mbox{$\Delta E$}}
\def\bp{\mbox{$B^+$}}
\def\bm{\mbox{$B^-$}}
\def\bpm{\mbox{$B^{\pm}$}}

\def\ks{\mbox{$K^0_S$}}
\def\pippim{\mbox{$\pi^+\pi^-$}}
\def\kpmpiz{\mbox{$K^{\pm}\pi^0$}}
\def\kspip{\mbox{$\ks\pi^+$}}
\def\kspim{\mbox{$\ks\pi^-$}}
\def\kspipm{\mbox{$\ks\pi^{\pm}$}}
\def\kzpip{\mbox{$K^0\pi^+$}}
\def\kzpim{\mbox{$\overline{K}{}^0\pi^-$}}
\def\KandKbar{\mbox{$K$\llap{\raise 0.8em\hbox to 1em{\tiny(---)\hss}}}}
\def\kzpipm{\mbox{\KandKbar${}^0\pi^{\pm}$}}
\def\dplus{\mbox{$D^+$}}
\def\dminus{\mbox{$D^-$}}
\def\dpm{\mbox{$D^{\pm}$}}
\def\dz{\mbox{$D$\llap{\raise 0.8em\hbox to 1em{\tiny(---)\hss}}${}^0$}}
\def\kpmpimp{\mbox{$K^{\pm}\pi^{\mp}$}}
\def\kmppipm{\mbox{$K^{\mp}\pi^{\pm}$}}

\def\dzpipm{\mbox{$\dz(\to\kpmpimp)\pi^{\pm}$}}

\def\acpvalold{\mbox{$0.46\pm 0.15\pm 0.02$}}
\def\nbbvalue{\mbox{$85.0 \pm 0.5$}}
\def\nbbv{\mbox{$85$}}
\def\nkspimvalue{\mbox{$119.1^{\ +13.8}_{\ -13.1}$}}
\def\nkspipvalue{\mbox{$104.4^{\ +13.2}_{\ -12.5}$}}
\def\acpvaluea{\mbox{$0.07^{\ +0.09}_{\ -0.08}$}}
\def\acpvalueb{\acpvaluea\mbox{$^{\ +0.01}_{\ -0.03}$}}
\def\acpinterval{\mbox{$-0.10 < \acp(\kzpipm) < 0.22$}}

\def\mySpecialText{DRAFT version 2.8 \today}
\def\myspecial#1{}                   

\begin{document}

\myspecial{!userdict begin /bop-hook{gsave 300 50 translate 5 rotate
    /Times-Roman findfont 18 scalefont setfont
    0 0 moveto 0.70 setgray
    (\mySpecialText)
    show grestore}def end}

\vspace*{-3\baselineskip}
\resizebox{!}{3cm}{\includegraphics{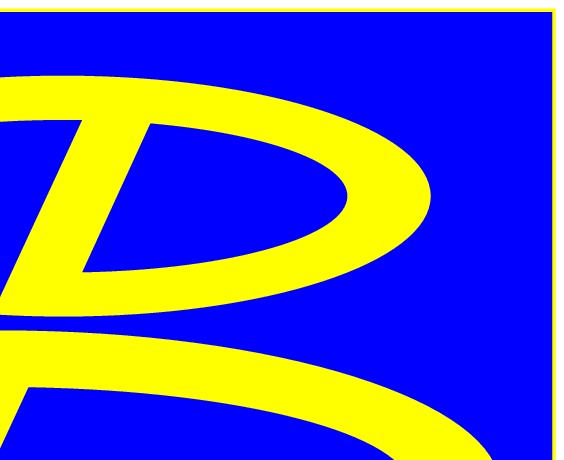}}
\preprint{\tighten\vbox{\hbox{\hfil KEK preprint 2003-7}}}
\preprint{\tighten\vbox{\hbox{\hfil Belle preprint 2003-4}}}

\vspace {1cm}

\title{ \Large \boldmath Improved Measurement of the Partial-Rate $CP$
	Asymmetry\\in $\bp\to\kzpip$ and $\bm\to\kzpim$ Decays }

\affiliation{Aomori University, Aomori}
\affiliation{Budker Institute of Nuclear Physics, Novosibirsk}
\affiliation{Chiba University, Chiba}
\affiliation{Chuo University, Tokyo}
\affiliation{University of Cincinnati, Cincinnati, Ohio 45221}
\affiliation{University of Frankfurt, Frankfurt}
\affiliation{Gyeongsang National University, Chinju}
\affiliation{University of Hawaii, Honolulu, Hawaii 96822}
\affiliation{High Energy Accelerator Research Organization (KEK), Tsukuba}
\affiliation{Hiroshima Institute of Technology, Hiroshima}
\affiliation{Institute of High Energy Physics, Chinese Academy of Sciences, Beijing}
\affiliation{Institute of High Energy Physics, Vienna}
\affiliation{Institute for Theoretical and Experimental Physics, Moscow}
\affiliation{J. Stefan Institute, Ljubljana}
\affiliation{Kanagawa University, Yokohama}
\affiliation{Korea University, Seoul}
\affiliation{Kyoto University, Kyoto}
\affiliation{Kyungpook National University, Taegu}
\affiliation{Institut de Physique des Hautes \'Energies, Universit\'e de Lausanne, Lausanne}
\affiliation{University of Ljubljana, Ljubljana}
\affiliation{University of Maribor, Maribor}
\affiliation{University of Melbourne, Victoria}
\affiliation{Nagoya University, Nagoya}
\affiliation{Nara Women's University, Nara}
\affiliation{National Kaohsiung Normal University, Kaohsiung}
\affiliation{National Lien-Ho Institute of Technology, Miao Li}
\affiliation{Department of Physics, National Taiwan University, Taipei}
\affiliation{H. Niewodniczanski Institute of Nuclear Physics, Krakow}
\affiliation{Nihon Dental College, Niigata}
\affiliation{Niigata University, Niigata}
\affiliation{Osaka City University, Osaka}
\affiliation{Osaka University, Osaka}
\affiliation{Panjab University, Chandigarh}
\affiliation{Peking University, Beijing}
\affiliation{Princeton University, Princeton, New Jersey 08545}
\affiliation{RIKEN BNL Research Center, Upton, New York 11973}
\affiliation{Saga University, Saga}
\affiliation{University of Science and Technology of China, Hefei}
\affiliation{Seoul National University, Seoul}
\affiliation{Sungkyunkwan University, Suwon}
\affiliation{University of Sydney, Sydney NSW}
\affiliation{Tata Institute of Fundamental Research, Bombay}
\affiliation{Toho University, Funabashi}
\affiliation{Tohoku Gakuin University, Tagajo}
\affiliation{Tohoku University, Sendai}
\affiliation{Department of Physics, University of Tokyo, Tokyo}
\affiliation{Tokyo Institute of Technology, Tokyo}
\affiliation{Tokyo Metropolitan University, Tokyo}
\affiliation{Tokyo University of Agriculture and Technology, Tokyo}
\affiliation{Toyama National College of Maritime Technology, Toyama}
\affiliation{University of Tsukuba, Tsukuba}
\affiliation{Utkal University, Bhubaneswer}
\affiliation{Virginia Polytechnic Institute and State University, Blacksburg, Virginia 24061}
\affiliation{Yokkaichi University, Yokkaichi}
\affiliation{Yonsei University, Seoul}
  \author{Y.~Unno}\affiliation{Chiba University, Chiba} 
  \author{K.~Suzuki}\affiliation{High Energy Accelerator Research Organization (KEK), Tsukuba} 
  \author{K.~Abe}\affiliation{High Energy Accelerator Research Organization (KEK), Tsukuba} 
  \author{K.~Abe}\affiliation{Tohoku Gakuin University, Tagajo} 
  \author{T.~Abe}\affiliation{Tohoku University, Sendai} 
  \author{I.~Adachi}\affiliation{High Energy Accelerator Research Organization (KEK), Tsukuba} 
  \author{H.~Aihara}\affiliation{Department of Physics, University of Tokyo, Tokyo} 
  \author{M.~Akatsu}\affiliation{Nagoya University, Nagoya} 
  \author{Y.~Asano}\affiliation{University of Tsukuba, Tsukuba} 
  \author{T.~Aso}\affiliation{Toyama National College of Maritime Technology, Toyama} 
% ...
\collaboration{The Belle Collaboration}

\begin{abstract}
We report an improved measurement of the partial-rate $CP$ asymmetry
in $\bpm\to\kzpipm$ decays. The analysis is based on a data sample of
$\nbbv$ million $\bb$ pairs collected at the $\Upsilon$(4S) resonance
with the Belle detector at the KEKB $e^+e^-$ storage ring.
We measure $\acp(\kzpipm) = \acpvalueb$, where the first and second
errors are statistical and systematic, respectively;
the corresponding 90\% confidence-level interval is $\acpinterval$.
\end{abstract}


\maketitle

In the Kobayashi-Maskawa (KM) model~\cite{km}, $CP$ violation arises
from a complex phase in the quark-mixing matrix of the weak interaction.
This idea is strongly supported by the observation of mixing-induced
$CP$ violation at the $B$-factories~\cite{phi1}.
Direct $CP$ violation (DCPV) is also expected in the KM scheme and
has been observed in the $K$ meson system~\cite{dcpv_k}.
However, DCPV has not yet been observed in the $B$ meson system.

Charmless hadronic $B$ decays can provide opportunities to observe 
DCPV~\cite{dcpv_b_theory,dcpv_b_belle,hh_prd,dcpv_b_cleo,dcpv_b_babar}.
Many of these decays include contributions from both $b\to u$ tree and
$b\to s$ penguin diagrams and the interference between these two
processes can produce a partial-rate $CP$ asymmetry:
\begin{eqnarray*}
 \acp &=& \frac{ \Gamma(\overline{B} \to \overline{f}) - \Gamma(B \to f) }
               { \Gamma(\overline{B} \to \overline{f}) + \Gamma(B \to f) }\\
      &=& \frac{ 2|A_T||A_P|\sin\delta\sin\phi }
               { |A_T|^2 + |A_P|^2 + 2|A_T||A_P|\cos\delta\cos\phi}.
\end{eqnarray*}
Here $\Gamma(B \to f)$ denotes the partial width of either a $B^0_d$
or $B^+$ meson decaying into a flavor-specific final state $f$ and
$\Gamma(\overline{B} \to \overline{f})$ represents that of the charge
conjugate decay;
$A_T$ and $A_P$ represent the tree and penguin amplitudes;
and $\delta$ and $\phi$ stand for the $CP$-conserving and $CP$-violating
relative phases, respectively, between $A_T$ and $A_P$.
In order to have a sizable ${\cal A}_{CP}$, both phase differences have
to be non-zero, i.e. $\delta \neq 0$ and $\phi\neq 0$, and the tree and
penguin amplitudes should be of comparable size ($|A_T| \sim |A_P|$).

The decay $\bpm\to\kzpipm$ is almost a pure $b\to s$ penguin process and,
thus, no sizable asymmetry is expected in the context of the Standard
Model (SM)~\cite{factorization,pqcd}.
However, our previously published result, based on an analysis of a 29
$\fb$ data sample, was $\acp(\kzpipm) = \acpvalold$~\cite{hh_prd}.
An asymmetry of this magnitude cannot be explained in the SM, even with
the inclusion of interference of the basic penguin amplitude with a large
$\bpm\to(\kpmpiz)_{\rm tree}\to\kzpipm$ re-scattering
process~\cite{rescattering}, and would be an indication of a new physics
contribution in the penguin loop~\cite{newphysics}.
It is important to verify whether the central value persists with improved
precision.

In this paper, we report an updated measurement of the partial-rate $CP$
asymmetry in $\bpm\to\kzpipm$ decays based on a 78 $\fb$ data sample
collected at the $\Upsilon$(4S) resonance, corresponding to $\nbbvalue$
million $\bb$ pairs, with the Belle detector~\cite{belle} at the KEKB
$e^+e^-$ storage ring~\cite{kekb}.
This is approximately three times as much data as the sample that was used
for the previous measurement and significantly improves the statistical
precision.
Throughout this paper, the partial-rate $CP$ asymmetry $\acp(\kzpipm)$ is
defined as
$$\acp(\kzpipm) \equiv
	\frac{ N(\kspim) - N(\kspip) }{ N(\kspim) + N(\kspip) },$$
where $N(\kspim)$ denotes the yield of $\bm\to\kspim$ decay and $N(\kspip)$
represents that of the charge conjugate mode.

The Belle detector is a large-solid-angle spectrometer that consists of
a three-layer silicon vertex detector (SVD), a 50-layer central drift
chamber (CDC), an array of threshold \v{C}erenkov counters with silica
aerogel radiators (ACC), time-of-flight scintillation counters (TOF),
and an electromagnetic calorimeter comprised of CsI(Tl) crystals (ECL)
located inside a superconducting solenoid coil that provides a 1.5 T
magnetic field.
An iron flux-return located outside of the coil is instrumented to detect
$K^0_L$ mesons and to identify muons (KLM).
A detailed description of the Belle detector can be found
elsewhere~\cite{belle}.

The analysis procedure is the same as described in Ref.~\cite{hh_prd}.
Candidate $\bpm$ mesons are reconstructed using high momentum
$\pi^{\pm}$ and $\ks$ mesons.
For candidate $\pi^{\pm}$ mesons, charged tracks are required to
originate from the interaction region based on their impact parameters.

In Belle, high momentum $\pi^{\pm}$ and $K^{\pm}$ mesons are distinguished
by their associated \v{C}erenkov light yield ($N_{\rm p.e.}$) in the ACC
and the ionization  energy loss ($dE/dx$) in the CDC.
These quantities are used to form a particle identification (PID) likelihood
ratio ${\cal R}_{\pi} = {\cal L}_{\pi} / ( {\cal L}_{\pi} + {\cal L}_K )$,
where ${\cal L}_{\pi}$ denotes the product of the individual likelihoods
of $N_{\rm p.e.}$ and $dE/dx$ for $\pi^{\pm}$ mesons;
${\cal L}_K$ is the product for $K^{\pm}$ mesons.
For the ${\cal R}_{\pi}$ requirement used in this analysis, $\pi^{\pm}$
mesons are identified with an efficiency of $91$\% and there is a $10$\%
$K^{\pm}$ misidentification rate.
The efficiency and fake rate are estimated by comparing the $\dz$ yields
in a sample of $D^{*\pm}$-tagged $\dz\to\kmppipm$ decays before and after
the application of the high momentum PID requirements.
A similar likelihood ratio that also includes the energy deposit in the ECL
is used to identify electrons; positively identified electrons are rejected.
Candidate $\ks$ mesons are reconstructed using pairs of oppositely
charged tracks that have an invariant mass ($m_{\pi\pi}$) in the range
$480 < m_{\pi\pi} < 516$ MeV/$c^2$. A candidate must have a displaced
vertex and flight direction consistent with a $\ks$ originating from the IP.

Signal candidates are identified using the beam-energy constrained mass
$\mbc = \sqrt{E_{\rm beam}^2 - p_B^{*2}}$ and the energy difference
 $\de = E_B^* - E_{\rm beam}$, where $E_{\rm beam}=5.29$~GeV and $p_B^*$
and $E_B^*$ are the momentum and energy of the reconstructed $B$ meson in
the $e^+e^-$ center-of-mass frame.

The dominant background comes from the $e^+e^-\to\qq$ ($q = u$, $d$, $s$, $c$)
continuum process;
backgrounds from $b\to c$ decays are negligible because the momenta of the
decay products are smaller than those of the signal $\ks$ and $\pi^{\pm}$.
We discriminate the signal from the $\qq$ background by the event topology. 
This is quantified by the Super-Fox-Wolfram ($SFW$) variable~\cite{hh_prd,fw}, 
which is formed from modified Fox-Wolfram moments that are combined using
a Fisher discriminant~\cite{fisher} into a single variable. 
The angle of the $B$-meson's flight direction with respect to the beam axis
($\theta_B$) provides additional discrimination.
A likelihood ratio
${\cal R}_s = {\cal L}_s / ({\cal L}_s + {\cal L}_{q\overline{q}})$
is calculated, where ${\cal L}_s$ (${\cal L}_{q\overline{q}}$)
denotes the product of the individual $SFW$ and $\theta_B$ likelihoods
for signal ($\qq$ background).
The probability density functions (PDFs) are derived from GEANT-based
Monte Carlo (MC) simulations~\cite{geant} for the signal and $\mbc$
sideband ($5.2 < \mbc < 5.26$ GeV/$c^2$) data for the $\qq$ background.
We make a requirement on ${\cal R}_s$ that eliminates 88\% of the $\qq$
background while retaining 73\% of the signal.

\begin{figure}[htb]
\centerline{
\epsfig{file=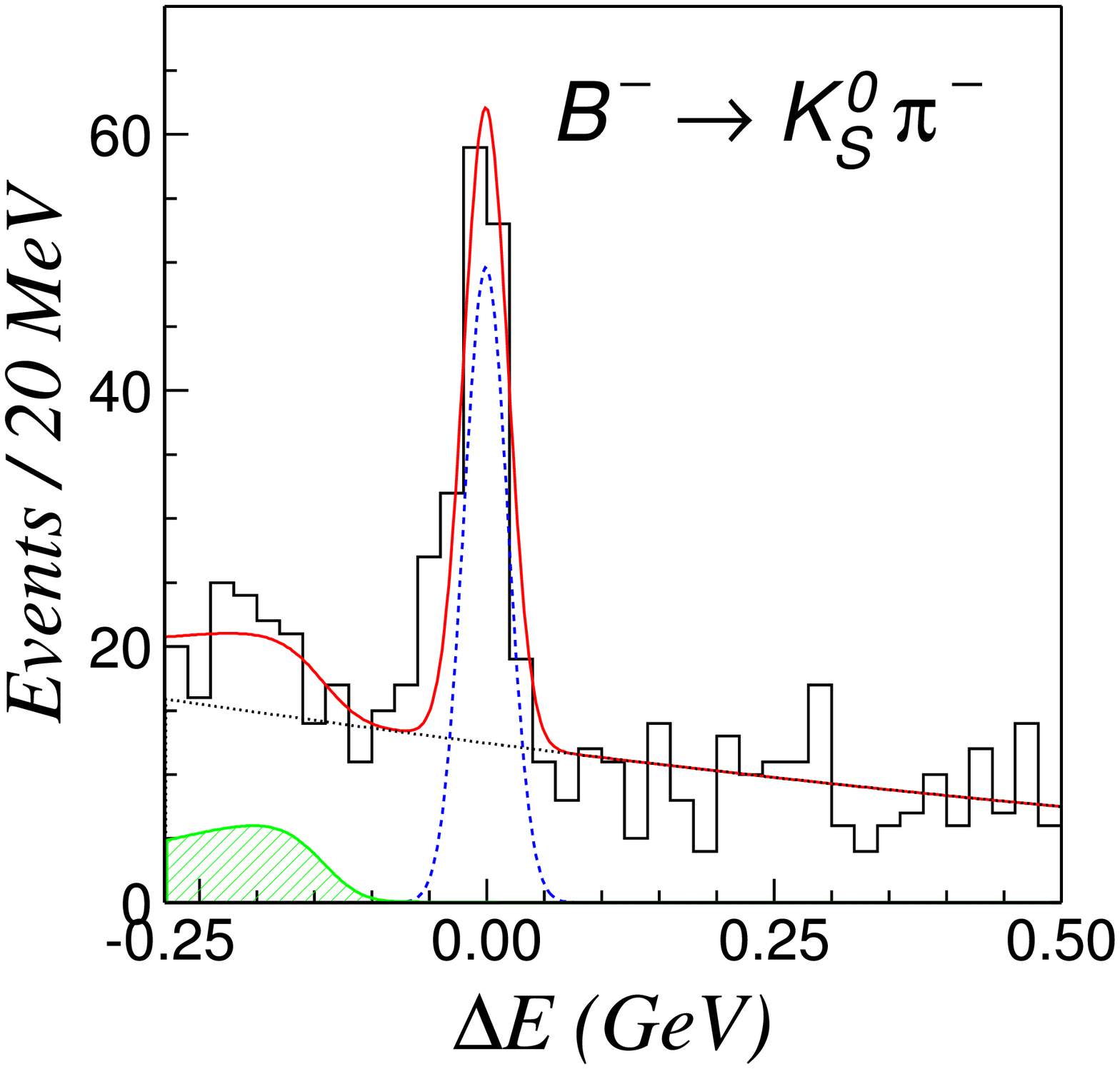,width=4.5cm}
\epsfig{file=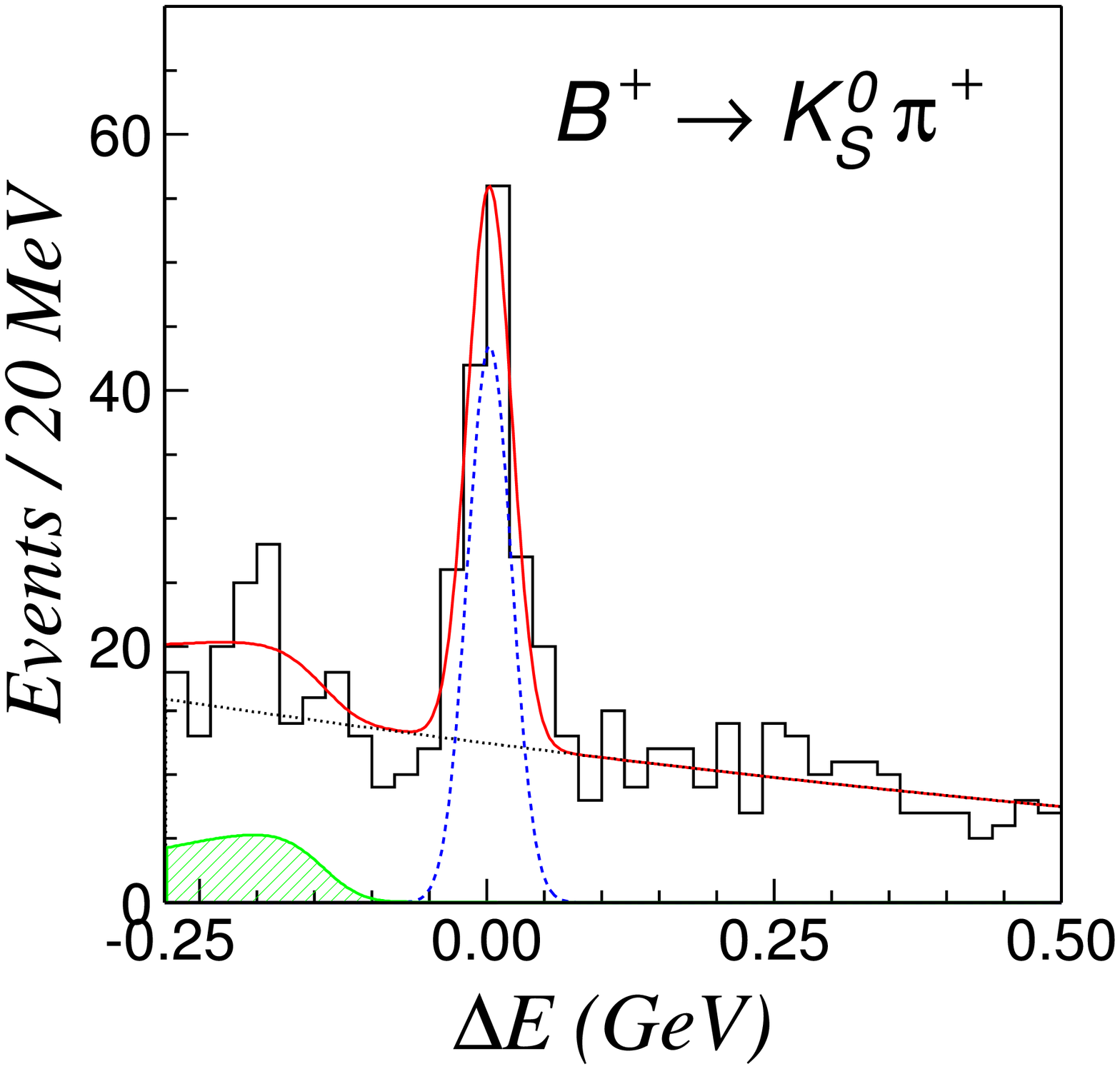,width=4.5cm}
}
\caption{
The $\de$ distributions for the $\bpm\to\kspipm$ candidates divided
into $\bm$ (left) and $\bp$ (right) samples. The fit results are
shown as the solid, dashed and dotted curves for the total, signal
and $\qq$ background, respectively; the hatched area indicates
the contribution from other charmless $B$ decays.
}
\label{fig:kspi_defit}
\end{figure}

Signal yields are extracted from the $\de$ distributions of events in
the $\mbc$ signal region ($5.271 < \mbc < 5.287$~GeV/$c^2$), separately
for the $\kspip$ and $\kspim$ final states.
The signal reconstruction efficiency~\cite{sig_eff} is estimated to be
12\% based on the MC. The $\de$ distributions are fitted using a binned
maximum likelihood fit with three components: the signal, $\qq$ background,
and other charmless $B$ decays, as shown in Fig.~\ref{fig:kspi_defit}.
The signal PDF is modeled with a Gaussian distribution taken from the
signal MC and calibrated using a $\bpm\to\dzpipm$ sample where a similar
reconstruction procedure is applied.
For the $\qq$ background, the PDF is modeled with a second-order polynomial
with a shape that is determined from the $\mbc$ sideband data.
For other charmless $B$ decays, the PDF is taken from a smoothed histogram
of a large MC sample.
(The enhancement in the lower $\de$ region is due to charmless $B$ decay
modes with an additional $\pi$ meson.)
Except for the signal peak positions, the same PDF shape parameters are
used for both $\bp$ and $\bm$ samples.
The signal peak positions are determined separately for the $\bp$ and
$\bm$ samples since a small systematic difference between the two samples
is observed. (This is discussed below.)
In the fit procedure, all of the PDF shape parameters are fixed and all
the normalizations are free parameters.
The signal yields are found to be $N(\kspip) = \nkspipvalue$ and
$N(\kspim) = \nkspimvalue$, and the partial-rate $CP$ asymmetry is
determined to be $\acp(\kzpipm) = \acpvaluea$.

\begin{figure}[htb]
\centerline{
\epsfig{file=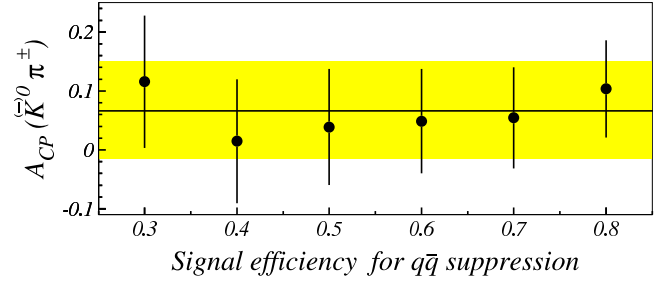,width=8.5cm}
}
\caption{
$\acp(\kzpipm)$ as a function of the signal efficiency of the $\qq$
suppression (${\cal R}_s$) selection. 
The horizontal line and hatched area indicate the $\acp(\kzpipm)$
value and its statistical error for the ${\cal R}_s$ requirement used
in the actual measurement.
Note that the statistical errors for the different data points are
strongly correlated.
}
\label{fig:comp_acp_lreff}
\end{figure}

The stability of $\acp(\kzpipm)$ as a function of the the selection
requirements is tested by varying the $\qq$ suppression requirement.
As shown in Fig.~\ref{fig:comp_acp_lreff}, the value of $\acp(\kzpipm)$
is stable when this requirement is changed.

\begin{figure}[htb]
\centerline{
\epsfig{file=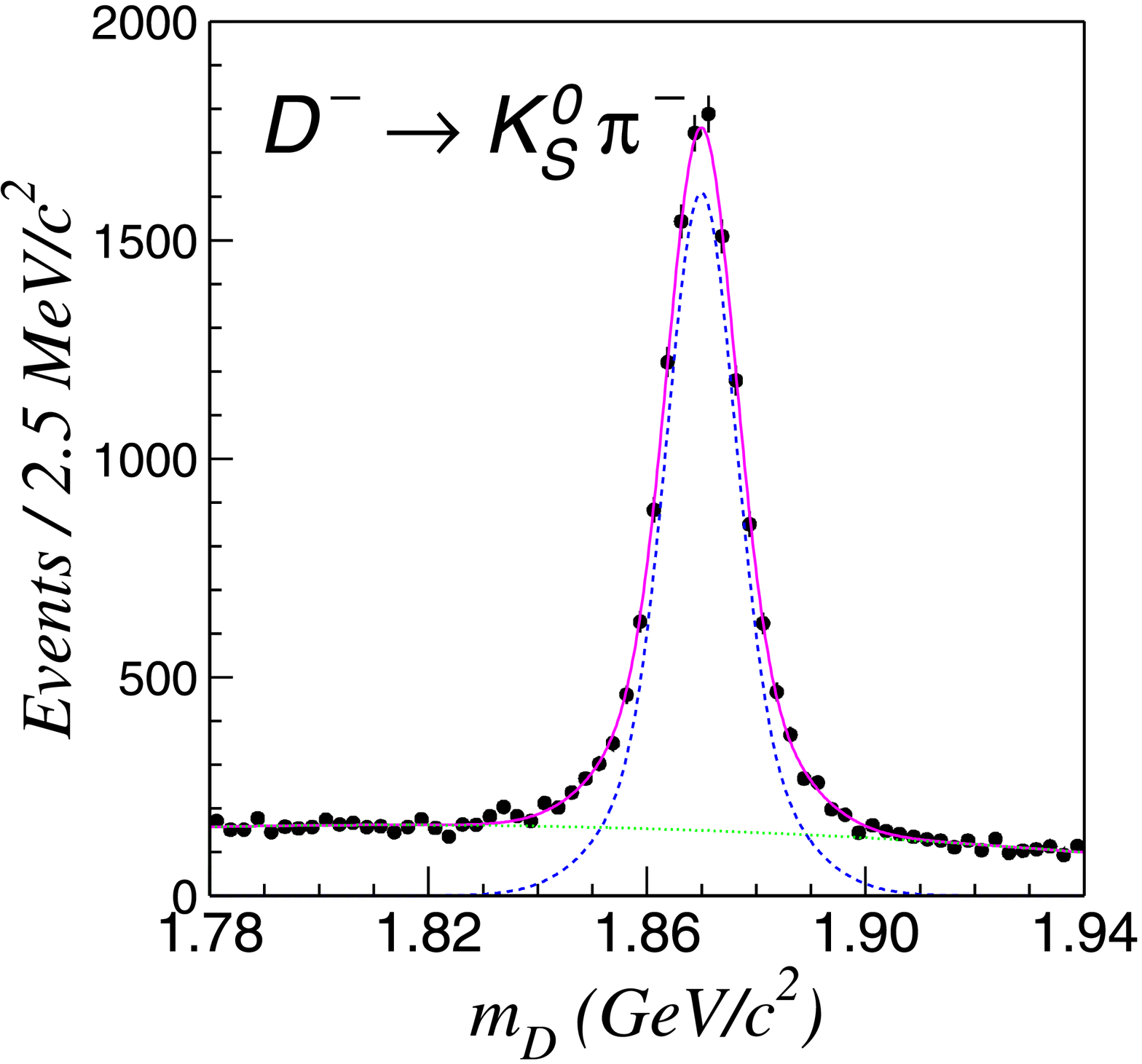,width=4.5cm}
\epsfig{file=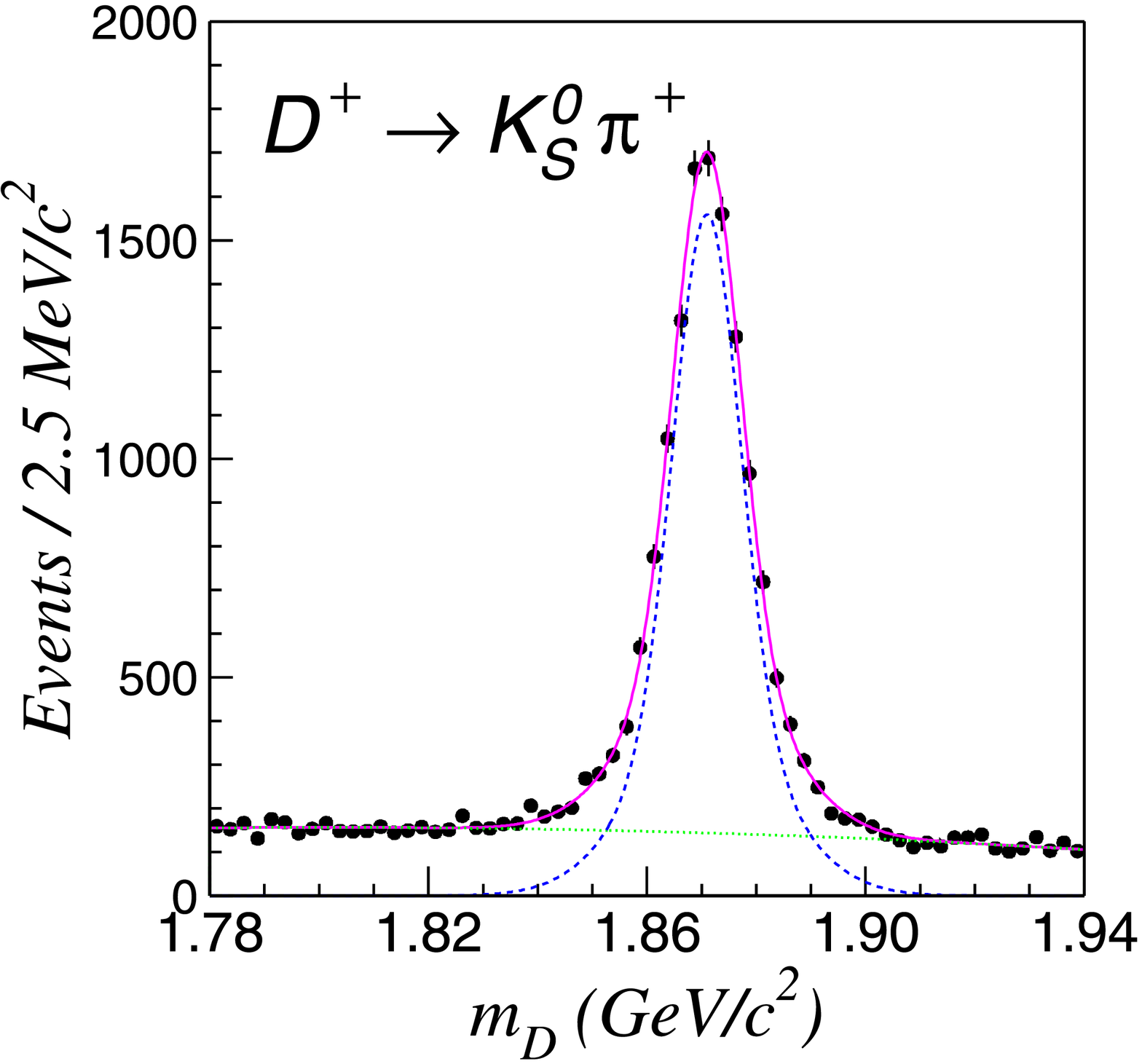,width=4.5cm}
}
\caption{
The mass spectra for the $\dpm\to\kspipm$ candidates separated into
$\dminus$ (left) and $\dplus$ (right) samples, where the kinematic
requirements and daughter particle reconstruction are the same as used
for the $\bpm\to\kspipm$ signal.
The fit results are  shown as the solid, dashed and dotted curves for
the total, $\dpm\to\kspipm$ and combinatorial background, respectively.
}
\label{fig:kspi_mdfit}
\end{figure}

Detector-based biases in $\kspipm$ reconstruction are investigated using
a sample of inclusive, high momentum continuum $\dpm\to\kspipm$ decays,
where the daughter particles are required to satisfy the same kinematic
requirements and reconstruction criteria as used for the signal.
Separate fits to the $D^{\pm}$ mass distributions, shown in
Fig.~\ref{fig:kspi_mdfit}, indicate that the signal mass resolutions for
the $\bp$ and $\bm$ samples are consistent, but there is a $1.0\pm 0.1$
MeV/$c^2$ difference in the mass peak positions.
This difference in the peak positions is caused by a difference between
the momentum measurement for high momentum negative and positive tracks
that is attributed to a residual detector misalignment.
After accounting for this difference in peak positions,
$\acp(\dpm\to\kspipm)$ is determined and listed in
Table~\ref{tab:systematics}.
Here the sign convention in the definition of $\acp(\dpm\to\kspipm)$
follows that of $\acp(\kzpipm)$.
The observed $(2.0\pm 0.8)\%$ asymmetry is treated as a possible bias,
and $-2.8\%$ is assigned as a systematic error in the $\acp(\kzpipm)$
measurement.

\begin{figure}[htb]
\centerline{
\epsfig{file=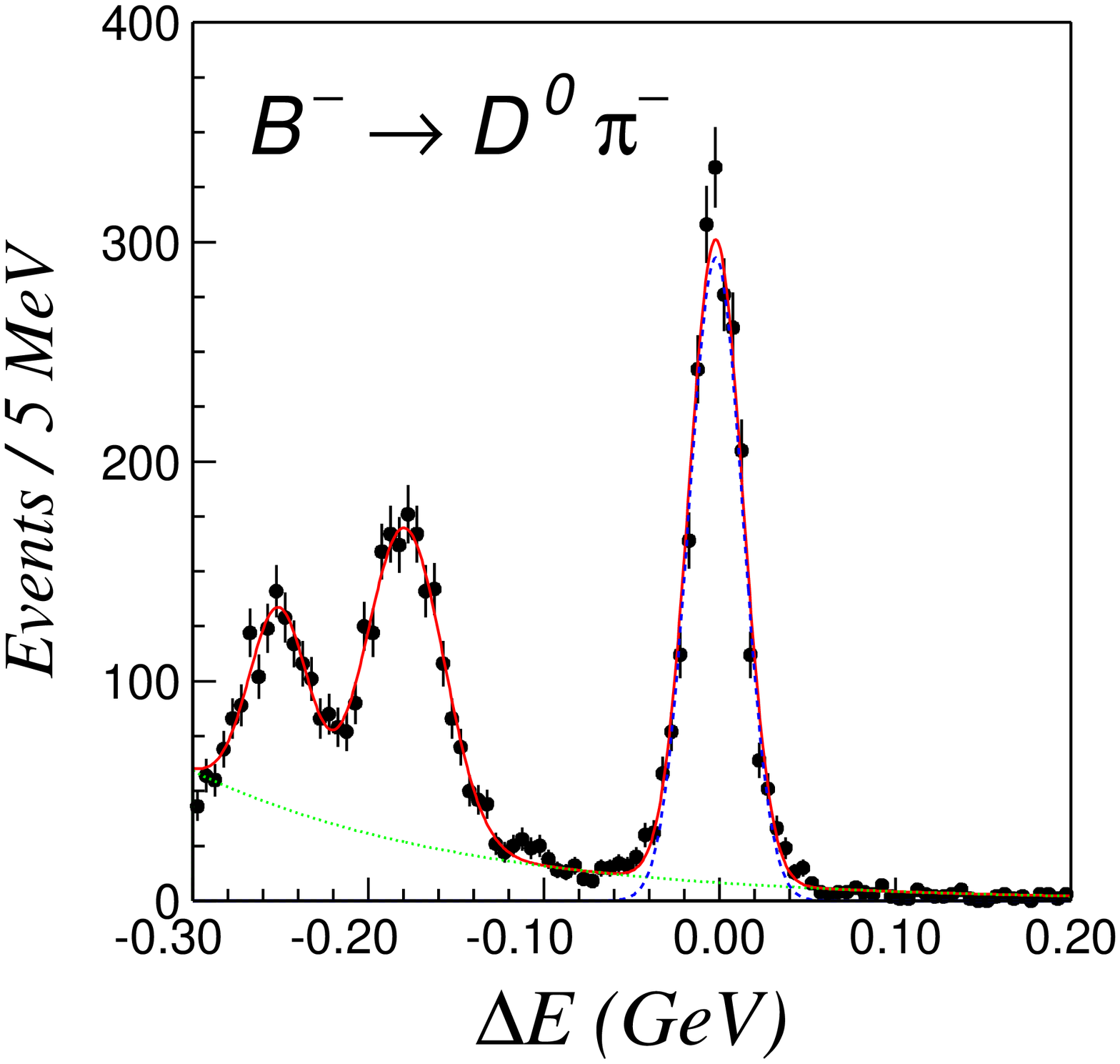,width=4.5cm}
\epsfig{file=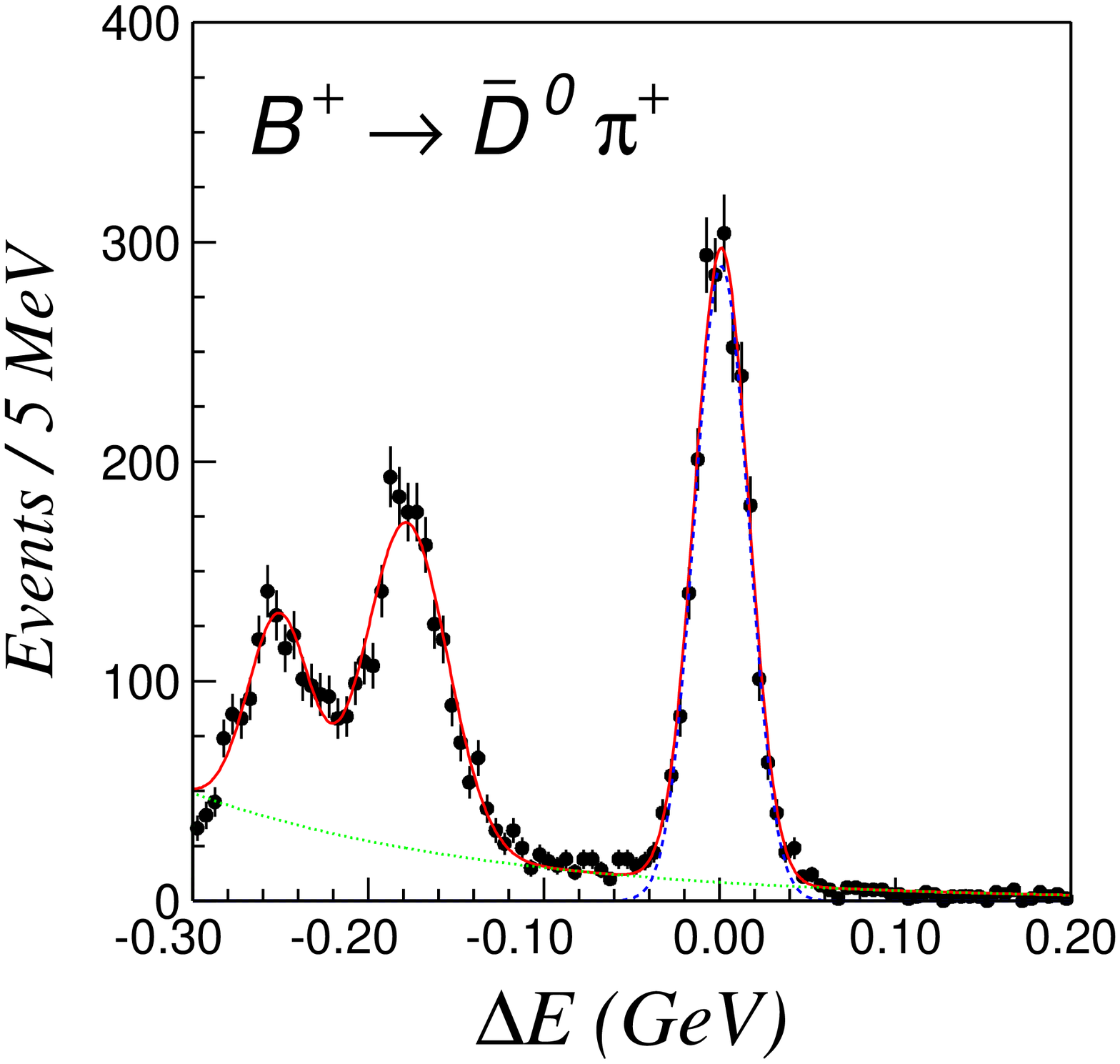,width=4.5cm}
}
\caption{
The $\de$ distributions for $\bpm\to\dzpipm$ candidates separately
for the $\bm$ (left) and $\bp$ (right) samples after application of
the entire reconstruction procedure other than that for the $\ks$.
The fit results are shown as the solid, dashed and dotted curves for
the total, $\bpm\to\dzpipm$ and combinatorial background, respectively.
The enhancement in the lower $\de$ region contains backgrounds from
$B\to D^*\pi^{\pm}$ and $D^0\rho$.
}
\label{fig:dpi_defit}
\end{figure}

Possible biases in the $B$ reconstruction are examined using a sample
of $\bpm\to\dzpipm$ decays where the entire reconstruction procedure,
except for the $\ks$ reconstruction, is applied.
Fits to the separate $\bp$ and $\bm$ $\de$ distributions, shown in 
Fig.~\ref{fig:dpi_defit}, confirm that the resolutions are consistent
between the two samples, while a $3.2\pm 0.5$~MeV difference in peak
positions, due to the same effect that was found in the $\dpm\to\kspipm$
sample, is observed.
After accounting for the difference in $\de$ peak positions,
$\acp(\bpm\to\dz\pi^{\pm})$ is determined and listed in
Table~\ref{tab:systematics}.
The absence of an asymmetry indicates there is no bias.
Biases in the high momentum PID and $\qq$ suppression are also examined
by removing each of them in the $\acp(\bpm\to\dz\pi^{\pm})$ measurement.
The results are given in Table~\ref{tab:systematics}.
No biases are observed.

Possible asymmetries in the detector response and reconstruction for the
$\qq$ background are checked using events in the $\mbc$ sideband region.
The application of the entire reconstruction procedure confirms that the
$\de$ shapes of the $\bp$ and $\bm$ samples are consistent, and no bias
is observed, as indicated in Table~\ref{tab:systematics}.
The absence of ${\cal R}_{\pi}$- and ${\cal R}_{s}$-related biases are
confirmed in the same manner as for the $\bpm\to\dzpipm$ sample.

In order to study the sensitivity to the signal and $\qq$ background
PDF shapes, each shape parameter is independently varied by its 1$\sigma$
error. In addition, the signal shape parameters are also estimated from
the actual $\bpm\to\kspipm$ samples by allowing them to be free parameters
in the fits.
The uncertainty in the contribution from other charmless $B$ decays is
estimated from the change in the asymmetry by fitting the region of
$\de > -0.1$ GeV without those decays.
The resulting relative changes in asymmetries are added in quadrature
giving the fitting systematics of $+0.014$ and $-0.006$.

\begin{figure}[htb]
\centerline{
\epsfig{file=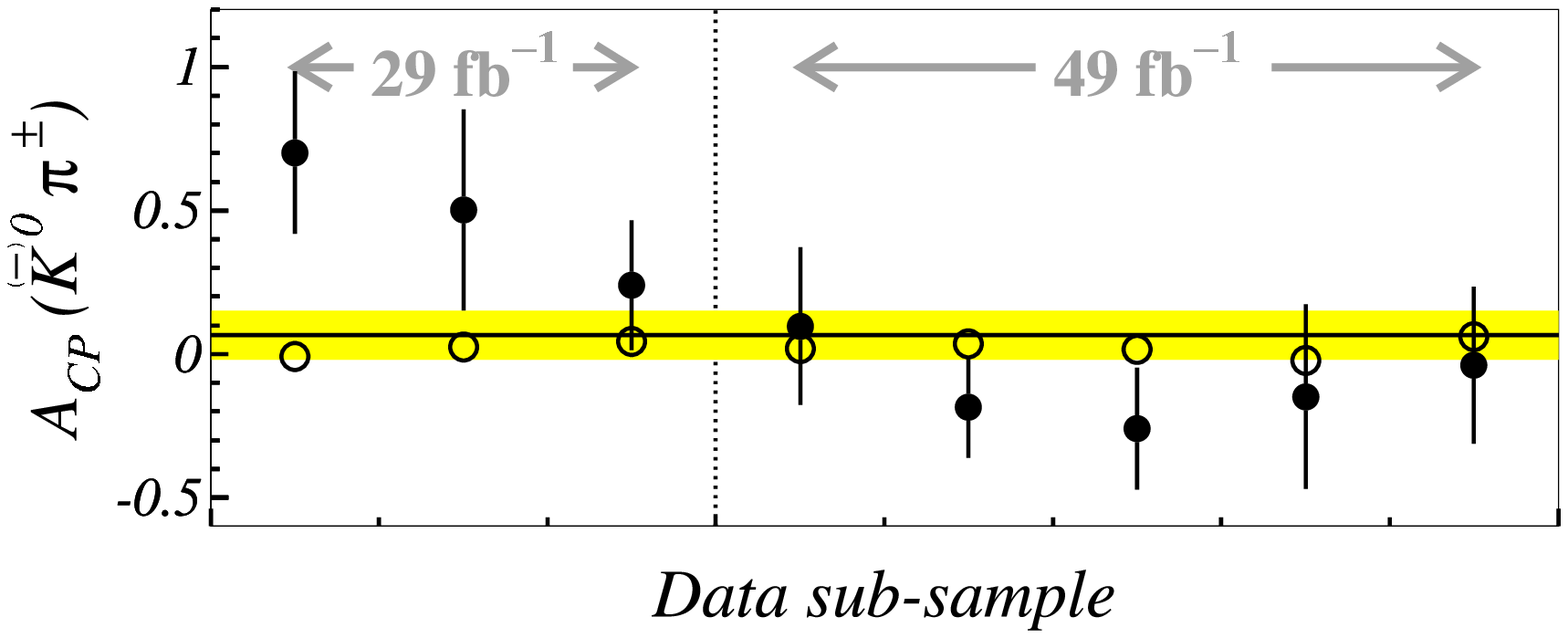,width=8.5cm}
}
\caption{
$\acp(\kzpipm)$ in each data sub-sample.
The horizontal line and hatched area show the central value and the
$1\sigma$ statistical error of the $\acp(\kzpipm)$ result reported here.
The solid points with the statistical error bars represent the
$\acp(\kzpipm)$ result obtained for each data sub-sample; the open points
show $\acp(\dpm\to\kspipm)$.
The sum of the three left-most points corresponds to the 29 $\fb$ data
sample used in our previous measurement.
}
\label{fig:comp_acp_dat}
\end{figure}

Because of the difference between the results presented here and the
sizable asymmetry in our previous measurement, the asymmetries of
different data sub-samples are examined.
Figure~\ref{fig:comp_acp_dat} shows $\acp(\kzpipm)$ for each data
sub-sample together with $\acp(\dpm\to\kspipm)$ as a reference.
The variation of $\acp(\kzpipm)$ is independent of that in
$\acp(\dpm\to\kspipm)$ and is consistent with statistical fluctuations.

The total systematic error in the $\acp(\kzpipm)$ is evaluated from
the quadratic sum of the $\kspipm$ reconstruction bias and $\de$
fitting systematics.
Finally, the asymmetry $$\acp(\kzpipm) = \acpvalueb$$ is obtained and
a 90\% confidence level interval $$\acpinterval,$$ is set, where Gaussian
statistics are assumed and the systematic error is added linearly.

In conclusion, we have measured the partial-rate $CP$ asymmetry in
$\bpm\to\kzpipm$ with $\nbbv$ million $\bb$ pairs collected on the
$\Upsilon$(4S) resonance at the Belle experiment.
The resulting $\acp(\kzpipm) = \acpvalueb$ is consistent with zero
at the current level of statistical precision.
The 90\% confidence level interval $\acpinterval$ is set, which is
consistent with other measurements~\cite{dcpv_b_cleo,dcpv_b_babar}.
This result has a statistical precision below 10\% and supersedes
our previous measurement~\cite{hh_prd}.
We do not observe a significant partial-rate $CP$ asymmetry in 
$\bpm\to\kzpipm$ and attribute the sizable $\acp(\kzpipm)$ found
previously in a much smaller data sample to a statistical fluctuation.

We are grateful to Y. Okada for useful discussions and comments.
We wish to thank the KEKB accelerator group for the excellent operation
of the KEKB accelerator.
We acknowledge support from the Ministry of Education, Culture, Sports,
Science, and Technology of Japan and the Japan Society for the Promotion
of Science;
the Australian Research Council and the Australian Department of Industry,
Science and Resources;
the National Science Foundation of China under contract No.~10175071;
the Department of Science and Technology of India;
the BK21 program of the Ministry of Education of Korea and the CHEP SRC
program of the Korea Science and Engineering Foundation;
the Polish State Committee for Scientific Research under contract
No.~2P03B 17017;
the Ministry of Science and Technology of the Russian Federation;
the Ministry of Education, Science and Sport of the Republic of Slovenia;
the National Science Council and the Ministry of Education of Taiwan;
and the U.S.\ Department of Energy.

\begin{table}[htb]
\caption{
Summary of the detector-based bias tests.
For tests other than those with the $\dpm\to\kspipm$ sample, $\acp$ 
values determined without the  high momentum PID (${\cal R}_{\pi}$)
and $\qq$ suppression (${\cal R}_s$)  requirements are also listed.
}
\label{tab:systematics}
\newcommand{\m}{\hphantom{$-$}}
\begin{ruledtabular}
\begin{tabular}{llc}
Samples
&			& $\acp$ [\%] \\
\hline
$\dpm\to\kspipm$
&			& \m$2.0\pm 0.8$ \\
$\bpm\to\dzpipm$		
& 			& \m$0.6\pm 1.7$ \\
& w/o ${\cal R}_{\pi}$	& \m$0.0\pm 1.5$ \\
& w/o ${\cal R}_s$	& \m$0.0\pm 1.4$ \\
$\bpm\to\kspipm$ $\mbc$ sideband data			
&			& \m$0.9\pm 1.3$ \\
& w/o ${\cal R}_{\pi}$	& \m$0.5\pm 0.9$ \\
& w/o ${\cal R}_s$	& \m$0.5\pm 0.4$ \\
\end{tabular}
\end{ruledtabular}
\end{table}


\end{document}